# Negative Energy:
# Why Interdisciplinary Physics Requires Multiple Ontologies

Benjamin W. Dreyfus, Benjamin D. Geller, Julia Gouvea, Vashti Sawtelle,
Chandra Turpen, and Edward F. Redish

*Department of Physics, University of Maryland, College Park MD 20742*

**Abstract.** Much recent work in physics education research has focused on ontological metaphors for energy, particularly the substance ontology and its pedagogical affordances. The concept of negative energy problematizes the substance ontology for energy, but in many instructional settings, the specific difficulties around negative energy are outweighed by the general advantages of the substance ontology. However, we claim that our interdisciplinary setting (a physics class that builds deep connections to biology and chemistry) leads to a different set of considerations and conclusions. In a course designed to draw interdisciplinary connections, the centrality of chemical bond energy in biology necessitates foregrounding negative energy from the beginning. We argue that the emphasis on negative energy requires a combination of substance and location ontologies. The location ontology enables energies both "above" and "below" zero. We present preliminary student data that illustrate difficulties in reasoning about negative energy, and the affordances of the location metaphor.

## INTRODUCTION

Energy is a central concept in physics, chemistry, and biology, and has been widely promoted [1] as a way to connect physics and chemistry to biology. Yet the concept of energy can be fractured for students along disciplinary lines.[2,3] In this paper we examine one consequence of building interdisciplinary coherence around energy: the effect on ontological metaphors for energy. We examine this question from a theoretical perspective in light of recent literature, and present initial student data that speak to the theoretical argument.

## ONTOLOGICAL METAPHORS FOR ENERGY

In recent years, a popular theme in the physics education research literature has been the use of ontological metaphors for energy: conceptual metaphors [4] that express "what kind of thing energy is".[5] Scherr et al. [5] identify three ontologies for energy found in student and expert discourse:
- *Substance:* energy as "stuff" contained **in** objects
- *Stimulus:* energy **acts on** objects
- *Vertical location*: objects are **at** higher or lower energies, by analogy to gravitational energy.

They note that "the stimulus metaphor is not common in expert physicists' discourse about energy," and likewise here we focus primarily on the substance and location metaphors, both of which are commonly used by expert physicists.

Scherr et al. go on to focus on the substance ontology, making the case for its pedagogical advantages and detailing how it can be used in instruction. Brewe [6] takes a similar approach, also focusing on the energy-as-substance metaphor as a central framework for the introductory physics curriculum. Lancor [7] examines the use of conceptual metaphors for energy in all three disciplines, and also focuses on the substance metaphor in its various manifestations.

All of these recent papers share a theoretical commitment to dynamic ontologies. As described by Gupta et al. [8], both novices and experts can reason across multiple ontological categories for the same concept in physics. This stands in contrast to the "static ontologies" view [9] that there is one correct ontological category corresponding to each entity, and misconceptions arise from ontological miscategorizations. Thus, when Scherr et al. and Brewe advocate for emphasizing the substance ontology in instruction, they are not claiming that the substance ontology is the "correct" ontology for energy; rather, their claims are based on the pedagogical affordances of this metaphor. These affordances include supporting the ideas that energy is conserved, can be located in objects, is transferred among objects [5], and is unitary (i.e., there is only one type of energy) [6] and/or can change form.[7]

However, they concede that one place where the substance metaphor encounters difficulties is the rep-resentation of negative energy, since a substance cannot ordinarily be negative. Scherr et al. resolve this concern with "the realization that potential energy depends not only on the system of mutually interacting objects but also on a reference point." In other words, it is possible to choose a reference point such that the potential energy of the system of interest is always positive, enabling the use of the substance metaphor. In Brewe's Modeling Instruction course, energy is first visually represented with pie charts, which emphasize conservation and unitarity. This representation breaks down when attempting to incorporate negative energy, and this provides the motivation to replace pie charts with bar charts [10], which can represent negative energy. However, it is less clear that bar charts embody the substance metaphor in the way that pie charts do, or how negative bars fit into the structure of this metaphor.

## INTERDISCIPLINARITY AND NEGATIVE ENERGY

Our research in this area is in the context of developing the NEXUS/Physics course [11], an introductory physics course[1] for undergraduate biology students that is focused on building interdisciplinary coherence between physics, biology, and chemistry. In a traditional introductory physics course, the energy unit focuses on mechanical energy: kinetic energy and macroscopically detectable potential energies. "Chemical energy" is treated as a black box (to account for where the missing mechanical energy went) if at all. [12] This approach comes up short for biology students, because most energy relevant in biological systems is chemical energy (i.e. energy changes associated with chemical bonds and chemical reactions).

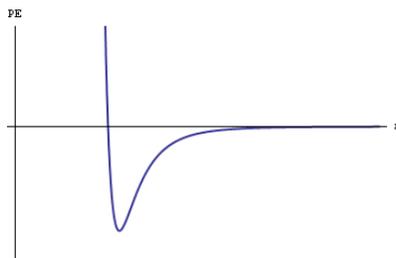

**FIGURE 1.** The Lennard-Jones potential, approximating the interaction between two atoms.

Therefore, chemical energy is a core component of the NEXUS/Physics course's treatment of energy, following other physics courses for the life sciences.[13] Electric forces and electric potential energy are moved up to the first semester and used to model (qualitatively) the potential for a system of two interacting atoms (Figure 1). This leads to a description of chemical bonds in terms of electric potential energy and other constructs that connect to the overall conceptual framework of physics.

The concept of negative energy is essential to this model of chemical bonds. When two atoms are bound, their energy is negative relative to the same atoms if they were unbound. Unlike models of gravitational potential energy that are common in introductory physics courses, the "zero" point of potential energy in this model is not arbitrary. Zero potential energy has a specific physical meaning here: the energy when the atoms are far enough apart that they are not interacting. Shifting the zero point below the strongest bond in the system to make all energies positive (in order to preserve the substance ontology) would mean that adding new molecules to the system (which have the capacity to form additional bonds) would require shifting the zero again, with no limit. Modeling bound atoms with negative energy contributes substantial conceptual clarity. Therefore, when chemical energy is a central piece of the overall energy picture, negative energy needs to be accessible from the beginning.

## MIXING THE ONTOLOGIES

While other authors operating in different instructional contexts have argued for the primary use of the energy-as-substance ontology, our student population and curricular goals lead us to a different cost-benefit analysis. Scherr et al. [5] are exploring these questions in the context of a professional development program for K-12 teachers, and Brewe's [6] Modeling Instruction course is for undergraduates from all the science and engineering majors. Neither context demands the same special concerns that are occasioned by our interdisciplinary context that attempts to form deep connections between physics and biology. The centrality of negative energy in our context means that an exclusive substance ontology for energy is untenable.

The energy-as-vertical-location metaphor is better suited for energies that can be positive or negative. While extending the substance ontology to negative quantities requires complicated maneuvering (e.g. defining a negative substance that cancels out when it combines with the positive substance), it is no more conceptually difficult to be at a location "below" zero than at a location "above" zero. The use of these two metaphors for negative numbers is explored extensively in the mathematics education literature [14], though not in the same language we use here. The location ontology for energy is also in common usage among expert physicists, such as in the potential well metaphor [15].

---

[1] See http://nexusphysics.umd.edu .

However, it is hard to imagine a comprehensive picture of energy that is based exclusively on the location ontology. The location metaphor succeeds at capturing some important aspects of energy: energy is a state function; energy can be positive or negative; changes in potential energy are more physically meaningful than the actual value of potential energy; intuitions based on gravitational potential energy about the relationship between energy and force (and embodied experience about up and down) can be applied to other non-gravitational energies. But there are other aspects where the location metaphor falls short: interactions and energy transfer among objects in a system; conservation.

Though neither the substance nor the location ontology is adequate on its own, combining the two addresses these shortcomings. And indeed, this particular combination is common among expert physicists. This is illustrated by the following classroom transcript from a physics professor teaching the NEXUS/Physics course. The use of the energy-as-substance metaphor is <u>underlined</u>, and the use of the energy-as-location metaphor is in **bold**. This coding excludes language (such as "get them back apart") that refers to the *spatial* location of the atoms, since that location is literal and is not a metaphor for energy.

> *If the two atoms are apart and form a bond, they **drop down to here** and <u>release that much energy</u>. And because that's **where they are, at that negative energy**, that's equal to <u>the energy you have to put in</u> to get them back apart. So it's just about **where you're going**, that when you're forming a bond, you're **dropping down**, and if you come in **at this energy** <u>you gotta get rid of this much</u>. But if **you're down here** and you want to **get back up to here**, <u>you gotta put in this much</u>.*

Here, the substance and location ontologies are combined in a productive way, and the professor fluidly moves between these metaphors within a single sentence. The mixed ontology is consistent: the energy of the system of atoms is described as a vertical location, and changes in the energy of the system are described as a substance (that enters or leaves the system). There is nothing extraordinary about this quotation; it illustrates a standard way that expert physicists talk about energy, especially in atomic and molecular contexts.

## STUDENT DATA

In the mathematics education context, Ball [14] writes that "comparing magnitudes becomes complicated. … Simultaneously understanding that -5 is, in one sense, more than -1 and, in another sense, less than -1 is at the heart of understanding negative numbers." Similar issues arise in physics, particularly in our interdisciplinary context. In most cases when we talk about negative energy, the "magnitude" is a distraction with no physical significance, since the zero point for potential energy is an arbitrary choice. In those cases, it is obvious that -5 is less than -1 (albeit not always obvious to students). However, in the context of chemical bonds, there is also a sense in which -5 is "more" than -1. A chemical bond with a deeper potential well, associated with a lower (more negative) potential energy, can also be described as a "stronger bond" or "more stable." In chemistry contexts, chemical binding energies are typically reported as positive quantities (absolute values).

We have observed this issue as a source of confusion among our students, as documented in the video data from the course, and we include a few examples here. In the NEXUS/Physics course, the students were working on a group problem-solving task that involved using energy bar charts to keep track of the energy in a biological process that included the formation and breaking of bonds. Phillip's[2] group drew all of the bars (including those representing the "chemical energy" associated with the bonds) as positive. When asked by the TA about this decision, Phillip responded:

> **Phillip:** *We said absolute value, like the magnitude of the energy.*
> **TA:** *Why did you decide to take the absolute value?*
> **Phillip:** *Because it doesn't really matter later on, because everything else, like this potential, whatever, it just matters where you put the zero.*

Phillip is avoiding negative energy by making all the energies positive, which is a valid move under some circumstances (possibly including the task his group is working on). However, he confuses two different methods of making negative quantities positive: translating all the potential energies by a constant amount (moving the zero), and taking the absolute value. While the former method preserves conservation of energy, the latter does not.

Working on a similar task in an interview, Anita took another approach. She drew negative bars for the potential energies associated with chemical bonds, and used the language of "increasing" to describe these bars becoming more negative. The bar chart representation may contribute to the idea that more negative energies are "more," since the size of the (negative) bar is larger. More generally, the substance metaphor may encourage thinking of more negative energies as "larger," while the location metaphor (associated with representations such as potential energy graphs) may encourage thinking of them as "lower." Depending on the context, one of these types of reasoning may be productive, while the other may be misleading.

---
[2] All names are pseudonyms.

Another well-documented issue in biology and chemistry education is the student difficulties around "energy stored in bonds."[3] The causes of this problem can be traced to multiple sources, but the inappropriate application of a substance ontology for energy may bear some responsibility. The substance ontology supports a metaphor in which a bond is a piñata containing "stuff," and the stuff (energy) is released when the bond is broken. Anita explained in class that she used to think about bonds this way: *"whenever chemistry taught us like exothermic, endothermic, … I always imagined like the breaking of the bonds has like these little molecules that float out."* She was using this metaphor *"until I drew … the potential energy diagram, and that's when I realized, to break it you have to put in energy."* In a followup interview, Anita explained her use of the potential energy graph:

> *What I imagine it is, to get it to break, you need to <u>put in energy</u>. So to **get up the hill**, you need to <u>input energy</u> … Say that you're **bicycling up the hill**. You <u>need energy to put it in</u>, that's what breaks the bond, but to bring them back together, it's <u>released</u>. So I just think of—when you're **falling down**, if you're **going down a hill with a bike**, you're not <u>putting in energy</u> to the pedals, but yet your pedals keep going so there's <u>energy released</u>.*

Anita's exclusive use of the substance metaphor led her to claim incorrectly that energy is released when bonds are broken. Switching to a mixed substance/location ontology helped her analyze this situation correctly.

Because of the potential well metaphor, Betsy actually finds it more intuitive to think of the bond potential energy as negative than as positive. In an interview, she compared the choice of zero in Figure 1 with an alternative choice of zero below the bottom of the well, while pointing to a potential energy diagram similar to Figure 1: *"If this is the ground [top of the well], then yeah I'm gonna want to roll down more. If this is my ground though [bottom of the well], like thinking about the fact that I'm so far up and that I'm gonna fall into a well above, it's just not as easy for me to grasp."* We may expect that students will generally have more difficulty with the idea that energy can be negative [16], but in this case, the vertical location metaphor encourages a picture in which thinking of those energies as "rolling down" below the ground is the most natural for some students.

## CONCLUSION

We have laid out the theoretical argument for the need for a mixed ontology for energy in an interdisciplinary physics setting. We have also provided some examples of student data that demonstrate both the challenges in reasoning about negative energy and examples of productive reasoning. In our future work, we intend to analyze the student data in greater detail to draw conclusions about how these theoretical considerations play out in practice with students.

## ACKNOWLEDGMENTS

The authors thank Ayush Gupta, Jessica Clark, Abigail Daane, and Ben Van Dusen for their helpful comments on an earlier draft of this paper, and the UMD Physics and Biology Education Research Groups. This work is supported by the NSF Graduate Research Fellowship (DGE 0750616), NSF-TUES DUE 11-22818, and the HHMI NEXUS grant.